# LAGOVirtual
# A Collaborative Environment for the
# Large Aperture GRB Observatory

R. Camacho[1], R. Chacón[1], G. Díaz[1], C. Guada[2], V. Hamar[1], H. Hoeger[1 3]
A. Melfo[2], L. A. Núñez[1,2], Y. Pérez[2], C. Quintero[2], M. Rosales[2], R. Torrens[1],
and the LAGO Collaboration

[1] *Centro Nacional de Cálculo Científico Universidad de Los Andes*
*Corporación Parque Tecnológico de Mérida, Mérida 5101*
*{reinac, chaconreinaldo, vanessa, gilberto, torrens}@ ula.ve*
[2] *Centro de Física Fundamental, Departamento de Física, Facultad de Ciencias,*
*Universidad de Los Andes, Mérida 5101, Venezuela*
*{carlosg, melfo, nunez, quinteroc, misael}@ ula.ve*
[3] *Centro de Simulación y Modelado (CeSiMo), Facultad de Ingeniería,*
*Universidad de Los Andes, Mérida 5101, Venezuela*
*hhoeger@ula.ve*

## Abstract

We present the LAGOVirtual Project: an ongoing project to develop platform to collaborate in the Large Aperture GRB Observatory (LAGO). This continental-wide observatory is devised to detect high energy (around 100 GeV) component of Gamma Ray Bursts, by using the single particle technique in arrays of Water Cherenkov Detectors (WCD) at high mountain sites (Chacaltaya, Bolivia, 5300 m a.s.l., Pico Espejo, Venezuela, 4750 m a.s.l., Sierra Negra, Mexico, 4650 m a.s.l). This platform will allow LAGO collaboration to share data, and computer resources through its different sites. This environment has the possibility to generate synthetic data by simulating the showers through AIRES application and to store/preserve distributed data files collected by the WCD at the LAGO sites. The present article concerns the implementation of a prototype of LAGO-DR adapting DSpace, with a hierarchical structure (i.e. country, institution, followed by collections that contain the metadata and data files), for the captured/simulated data. This structure was generated by using the community, sub-community, collection, item model; available at the DSpace software. Each member institution-country of the project has the appropriate permissions on the system to publish information (descriptive metadata and associated data files). The platform can also associate multiple files to each item of data (data from the instruments, graphics, postprocessed-data, etc.).

## 1. LAGO Experiment

Gamma Ray Burst are characterized by a sudden emission of electromagnetic radiation (X/γ) during a short period of time (0.1 - 100 s). They occur at an average rate of a few events per day. Short GRBs (<2*s*) are thought to be generated by the gravitational coalescence of two compact objects (neutron stars or black holes) and long duration GRBs (lGRB), usually associated with the core collapse (*collapsar*) of a massive star.

Since photons coming from GRBs cannot penetrate easily the atmosphere, they started to be registered by space telescopes (HETE, INTEGRAL, Swift and GLAST, now Fermi Gamma-Ray Space Telescope) but as the photon energies increase, the photon flux decreases as a power law and satellites become impractical. However, with inexpensive ground-based experiments of large area, it is possible to detect relativistic secondary particles induced by the interaction of GeV/TeV gamma-ray photons with the molecules of the upper atmosphere. Water Cherenkov Detectors (WCD) emerge as efficient instruments for detecting GRBs in the 10 GeV - 1 TeV energy range by using the single particle technique [1, 2]. There are several ground-based experiments around the world searching for GRBs emision (Chacaltaya at 5200 m a.s.l. in Bolivia (INCA); Argo at 4300 m a.s.l. in Tibet; Milagro at 2650 m a.s.l. in New Mexico; the Pierre Auger Observatory at 1400 m a.s.l. in Malargüe, Argentina and Large Aperture GRB Observatory (LAGO) with several sites at high altitude (above 4500 m a.s.l.) in Mexico, Bolivia and Venezuela, with planned extension to Colombia, Guatemala, the Himalaya and Peru).

Naturally, researchers of this distributed community need to interact among them and to share data and computational resources in a continent-wide experiment. In the next section we will describe the project of `LAGOVirtual` for this research community.

## 2. `LAGOVirtual`

`LAGOVirtual` is an ongoing project devised to develop a Virtual Research Environment [3], i.e. a set of online tools to enhance the possibility for detecting GRB, to work in solar physics or any other application of the data collected by the WCD.

This environment allows researchers from different LAGO sites to work collectively accessing tools for simulation particle showers, remote access to the electronic of the detectors and to shared simulated and/or measured sets of data.

As it can be appreciated from the diagram in Figure 1, `LAGOVirtual` will have four modules:
- Remote access to the detectors, which will allow the user to modify the Photomultiplier (PTM) gain and baseline and the input channel for the Photomultiplier.
- Access to the AIRES, (A system for air shower simulations [4]) simulation environment and ROOT data analysis tool [5].
- Access to distributed data files.
- Access to distributed document files.

The present article reports the characteristic and the data structure of a prototype for the module of distributed repositories of data files. The next section will illustrate what are data repositories and the crucial importance that they will have for the curation of scientific data and the production of new knowledge.

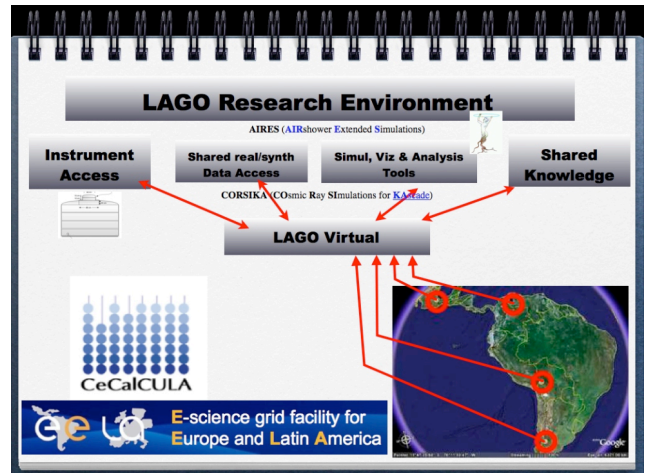

.

## 3. Data Repositories

For almost two decades electronic high energy physics preprints have been submitted to repositories, stored, indexed, retrieved, and shared. This initiative started the Open Access movement which now is building a network of Institutional Repositories (IR) worldwide. Although the importance of preserving the scientific data has been recommended long time ago [6], every year the data from the HEP experiments are lost, forgotten or kept for the restricted specific community of the collaboration[7]. Nowadays the cost, the sophistication and the rapid advancement of new experiments makes it essential that previous results remain accessible for accountability, re-analysis, and training of future generations. These needs have moved forward the Open Data Movements [8, 9, 10, 11].

At the moment, repositories are primarily designed for preprints but technically those documents could be linked to datasets hosted by the same repository or by others elsewhere. Having the data linked to the document or preserved in a repository students/researchers/groups outside the experimental collaboration could gain advantage from increased access to data.

Today there are some examples of data storage success in the astronomy, earth sciences, chemistry and biology fields. The Virtual Observatory, the National Space Science Data Center, the CombeChem, the eBank at the UK and European Bioinformatics Institute are emerging examples of this trend to link data to the information and knowledge. Future high energy physics data storage projects should seek lessons and advice from the staff working on these successful databases. Although the datasets for astronomy, chemistry, and biology tend to be less complex than those produced by typical experiments at particle accelerators, some issues are similar, such as questions of access, metadata definitions, and methods of software storage[7]. There are also some recent examples of environmental data en Latinamerica [12, 13].

In a data repository it's necessary to describe in some way all the documents and files inside it, otherwise it would be difficult to search and query, in this context the concept of metadata starts to play an important role. Metadata is data about data, it provides information about, or documentation of, other data managed within an application or environment. The metadata used depends of the kind of information in the repository, but there is a huge effort to create standard models of metadata in some fields of knowledge. This will make easier the interchange of documents and data fields between different repositories and systems.

## 4. LAGO Data Repositories

The LAGO Data Repositories (LAGO-DR) allows the collaboration to share data captured by the WCD and/or generated by simulations. Researchers at each institution-country can self archive data files generated by the instruments/simulations and they can also create automatic entries and files for bulk data load.

### 4.1. Data and metadata structure

We have developed a prototype of the data repository for the LAGO collaboration adapting the system Dspace ( http://www.dspace.org/ ), an open source software that enables open sharing of many types of content, generally used for document institutional repositories. Based on the community, sub-community, collection, item model; available at the DSpace software, we established a hierarchical structure starting with country who belong to the collaboration (community), institution (sub-community) and finally collections for different data types, to which the items (data and documents files) are associated.

So far, the data for LAGO are classified mainly into three types: instrument calibration data, data sets captured by the WCD instruments and data simulated by the AIRES application. In the future, we want the members of the collaboration to use this repository also to preserve papers, thesis, Labs Notes and/or other kind of documents related with the project.

Each data file (Dspace's items) is described by a metadata set specifically adapted to LAGO. The existence and implementation of a scientific metadata standard model will allow an uniform access to data for all the members of LAGO collaboration, the interoperability between scientific information systems and also will contribute to the data preservation and its usability in time. The metadata model we propose for LAGOvirtual is an adaptation of the model raised for the CCLRC (Council for the Central Laboratory of the Research Councils http://epubs.cclrc.ac.uk/bitstream/485/ ).

The Dublin Core metadata element set is a standard for cross-domain information resource description, elaborated and sponsored by DCMI (Dublin Core Metadata Initiative http://dublincore.org/ ). Implementations of Dublin Core typically make use of XML and are Resource Description Framework based http://www.w3.org/RDF/ . Dublin Core comprises fifteen metadata elements: Title, Subject, Description, Source, Language, Relation, Coverage, Creator, Publisher, Contributor, Rights, Date, Type, Format, and Identifier.

DSpace currently uses a qualified version of the Dublin Core standard. Taking advantage of the fact that CCLRC model's elements can be mapped to those of Dublin Core, we adapted Dspace adding new metadata useful to document the data files to be generated by the LAGO collaboration.

### 4.2. `LAGOvirtual` user interface

`LAGOvirtual` allows open access to its contents; it means that anyone can navigate and search within the repository. It's possible to make a list of the information available in the repository using different criteria such as: by country, by person responsible of data, by name of the data file and/or by type of file. Also in the central column there are forms and links to recover data organized by country and by institution. These facilities make easier and userfriendly the search of a given data file by the members of the collaboration or any other person interested in data deposited there.

However, only the members of the project have the permission to public new information. In order to deposit new information the user can click the option "Mi LAGOVirtual" located in the right menu, then he has to enter in the selected collection and click on "Enviar un item en esta colección". He will fill the form, which asks for information related with the person responsible of data taking and his contact information, name of the data file, date and hour of start and end data capture, which calibration file of the PMT are associated with the data file, the resources used, any problems during the capture process, temperature and voltage of the PMT. `LAGOvirtual` allow to associate different file (data files, calibration files, graphics, post-processed data, etc.) to the same item.

In Fig. 2. we show several screen captures of the prototype´s Web interface. From left to right, the main page of the Web portal; the screen with the list of countries members of the project; and at last a view of an item containing different data and image files.

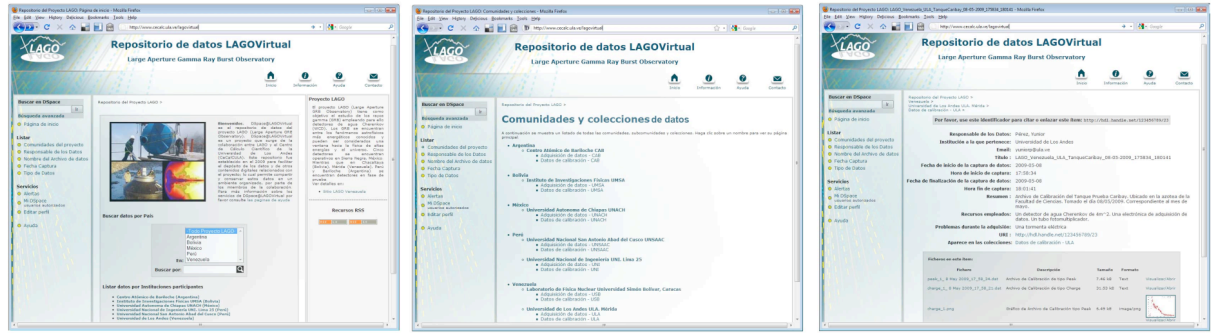

**Fig. 2. `LAGOVirtual` Web interface**

It is possible to associate a Creative Commons <u>http://creativecommons.org/</u> license to each file in the repository. The Creative Commons license (CC) is inspired in the GPL license (General Public License) of the Free Software Foundation and they provide a legal model to facilitate the distribution and use of contents. There are a series of licenses Creative Commons with different configurations, like the original right author to quote his work, reproduce it, offer it to the public and with different restrictions like to allow his commercial use or respect the copyright.

It is possible to subscribe to a system of automatic email notification, in case that new files were deposited in a particular collection. At the end of each item published appears the option "recomendar este artículo", in this way the user can recommend content to anyone linked or not, to the project.

`LAGOvirtual` includes a module of statistics (left side of the repository), which allows the members of the collaboration to know how many persons visit the repository, how many files have been downloaded in a given period of time, and which files are the more downloaded and/or consulted.

`LAGOvirtual` enables several mechanisms to share the data and metadata with other sytems. Dspace can expose the data and metadata through the Open Archives Initiatives Protocol for Metadata Harvesting (OAI-PMH <u>http://www.openarchives.org/pmh/</u>). This protocol is used by external systems to collect the data and metadata and create services of aggregated value like meta-searchers. Also offers the data through Really Simple Syndication channels, RSS <u>http://en.wikipedia.org/wiki/</u>, available at all levels in the structure of the repository (communities and collections). These channels are a simple mechanism to show contents recently submitted.

## 3. Summary and conclusions

We have described the data structure, the user interface and some of the functions of a prototype of the data repository for the LAGO collaboration. With this platform the collaboration can share data coming from the instrument calibration, data sets captured by the WCD instruments and data simulated by the AIRES application. The system not only allows the user to preserve the recorded data, but also to be notified when new data files are deposited. It also includes a statistics module for data mining in order to follow how many persons visit the repository, how many files have been downloaded at a given period of time, and which files are the more downloaded and/or consulted.

The data module `LAGOvirtual` allows open access to the data recorded for the entire collaboration, it means that anyone from inside the collaboration can navigate and search within the repository. Additionally, `LAGOvirtual` shares data and metadata with other systems and meta-searchers through the OAIPMH and RSS channel giving a significative visibility to the collaboration.

The `LAGOVirtual` data model will be a part of a VRE that allows researchers from different LAGO sites to work collectively accessing tools for simulation particle showers, remote access to the electronic of the detectors and to shared simulated and/or measured sets of data.

## Acknowledgments

This work was partially funded by the European Commission through the EELA 2 Project (**E**-Science grid facility for **E**urope and **L**atin **A**merica). We also gratefully acknowledge the financial support of the Consejo de Desarrollo Científico, Humanístico y Tecnológico-CDCHT-ULA under project C-1598-08-05-A

F-2002000426.

## References

[1] D. Allard. Detecting grbs with the Pierre Auger observatory using the single particle technique. *Nuclear Physics B (Proceedings Supplements)*, 165:110–115, 2007.

[2] D. Allard, I. Allekotte, C. Alvarez, H. Asorey, H. Barros, X. Bertou, O. Burgoa, M. Gomez Berisso, O. Martínez, and P. Miranda Loza. Use of Water-Cherenkov Detectors to detect Gamma ray bursts at the Large Aperture grb Observatory (LAGO). *Nuclear Inst. and Methods in Physics Research, A*, 595(1):70–72, 2008.

[3] A. Borda, J. Careless, M. Dimitrova, J. Fraser, M. Frey, P. Hubbard, S. Goldstein, C. Pung, M. Shoebridge, and N. Wiseman. Report of the working group on virtual research communities for the ost e-infrastructure steering group. Technical report, Office of Science and Technology, London UK, 2006.

[4] S. J. Sciutto. AIRES: A system for air shower simulations (version 2.2.0). Technical report, Universidad Nacional de la Plata, La Plata, Argentina, 1999.

[5] R. Brun and F. Rademakers. ROOT: An object oriented data analysis framework. *Nucl. Instrum. Meth.*, A389:81–86, 1997.

[6] J. Dozier, S. Alexander, M. Courain, J. A. Dutton, W. Emery, B. Gritton, R. Jenne, W. Kurth, D. Lide, B. K. Richard, and J. Warnow-Blewett. Preserving scientific data on our physical universe: A new strategy for archiving the nation's scientific information resources. Technical report, National Research Council, 1995.

[7] J. Yeomans. Archiving of Particle Physics Data and Results for Long-Term Access and Use. In A. Sissakian, G. Kozlov, & E. Kolganova, editor, *High Energy Physics: ICHEP '06*, pages 1201–+, 2007.

[8] P. Arzberger, A. Schroeder, A. Beaulieu, G. Bowker, K. Casey, L. Laaksonen, P. Uhlir, and P. Wouters. Promoting access to public research data for scientific,

economic, and social development. *Data Science Journal*, 3:135–152, Nov 2004.

[9]  L. Lyon. Dealing with data: Roles, rights, responsibilities and relationships consultancy report. Consultancy report, UKOLN, June 2007.

[10] M. Sabourin and B. Dumouchel. Canadian national consultation on access to scientific research data. *Data Science Journal*, 6(Open Data Issue):OD26–OD35, Jan 2007.

[11] G. Xu. Open access to scientific data: Promoting science and innovation. *Data Science Journal*, 6(Open Data Issue):OD21–OD25, Jan 2007.

[12] E. Barros, A. Laender, M. Gonçalves, and R. Cota. Transitioning from the ecological fieldwork to an online repository: a digital library solution and evaluation. *Int J Digit Libr*, 6:107–112, Jan 2007.

[13] H.Y. Contreras, Z. Méndez, R. Torrens, and L.A. Núñez. Desarrollo de la red bioclimática del estado Mérida, Venezuela: Estrategias de captura, manejo y preservación de datos ambientales. *Interciencia*, 33(11):795, 2008.